\documentstyle[aasms4]{article}
\makeatletter
  
  \@addtoreset{equation}{section}
\makeatother

\begin{document}

\title{RELATIVISTIC CORRECTIONS TO THE SUNYAEV-ZEL'DOVICH EFFECT FOR CLUSTERS OF GALAXIES. IV. ANALYTIC FITTING FORMULA FOR THE NUMERICAL RESULTS}

\author{SATOSHI NOZAWA}

\affil{Josai Junior College for Women, 1-1 Keyakidai, Sakado-shi, Saitama, \\
350-0295, Japan; snozawa@galaxy.josai.ac.jp}

\author{NAOKI ITOH AND YOUHEI KAWANA}

\affil{Department of Physics, Sophia University, 7-1 Kioi-cho, Chiyoda-ku, Tokyo, \\
102-8554, Japan; n\_itoh, y-kawana@hoffman.cc.sophia.ac.jp}

\centerline{AND}

\author{YASUHARU KOHYAMA}

\affil{Fuji Research Institute Corporation, 2-3 Kanda-Nishiki-cho, Chiyoda-ku, Tokyo, \\
101-8443, Japan; kohyama@star.fuji-ric.co.jp}

\begin{abstract}

  We present an accurate analytic fitting formula for the numerical results for the relativistic corrections to the thermal Sunyaev-Zel'dovich effect for clusters of galaxies.  The numerical results for the relativistic corrections have been obtained by numerical integration of the collision term of the Boltzmann equation.  The fitting is carried out for the ranges $0.02 \leq \theta_{e} \leq 0.05$ and $0 \leq X \leq 20$, where $\theta_{e} \equiv k_{B}T_{e}/m_{e}c^{2}$, $X \equiv \hbar \omega/k_{B}T_{0}$, $T_{e}$ is the electron temperature, $\omega$ is the angular frequency of the photon, and $T_{0}$ is the temperature of the cosmic microwave background radiation.  The accuracy of the fitting is generally better than 0.1\%.  The present analytic fitting formula will be useful for the analyses of the thermal Sunyaev-Zel'dovich effect for high-temperature galaxy clusters.

\end{abstract}


\section{INTRODUCTION}

  Compton scattering of the cosmic microwave background (CMB) radiation by hot intracluster gas --- the Sunyaev-Zel'dovich effect (Zel'dovich \& Sunyaev 1969; Sunyaev \& Zel'dovich 1972, 1980a, 1980b, 1981) --- provides a useful method to measure the Hubble constant $H_{0}$ (Gunn 1978; Silk \& White 1978; Birkinshaw 1979; Cavaliere, Danese, \& De Zotti 1979; Birkinshaw, Hughes, \& Arnaud 1991; Birkinshaw \& Hughes 1994; Myers et al. 1995; Herbig et al. 1995; Jones 1995; Markevitch et al. 1996; Holzapfel et al. 1997; Furuzawa et al. 1998).  The original Sunyaev-Zel'dovich formula has been derived from a kinetic equation for the photon distribution function taking into account the Compton scattering by electrons: the Kompaneets equation (Kompaneets 1957; Weymann 1965).  The original Kompaneets equation has been derived with a nonrelativistic approximation for the electron.  However, recent X-ray observations have revealed the existence of many high-temperature galaxy clusters (David et al. 1993; Arnaud et al. 1994; Markevitch et al. 1994; Markevitch et al. 1996; Holzapfel et al. 1997; Mushotzky \& Scharf 1997; Markevitch 1998).  In particular, Tucker et al. (1998) reported the discovery of a galaxy cluster with the electron temperature $k_{B} T_{e} = 17.4 \pm 2.5$ keV.  Rephaeli and his collaborator (Rephaeli 1995; Rephaeli \& Yankovitch 1997) have emphasized the need to take into account the relativistic corrections to the Sunyaev-Zel'dovich effect for clusters of galaxies.

  In recent years remarkable progress has been achieved in the theoretical studies of the relativistic corrections to the Sunyaev-Zel'dovich effects for clusters of galaxies.  Stebbins (1997) generalized the Kompaneets equation.  Itoh, Kohyama, \& Nozawa (1998) have adopted a relativistically covariant formalism to describe the Compton scattering process (Berestetskii, Lifshitz, \& Pitaevskii 1982; Buchler \& Yueh 1976), thereby obtaining higher-order relativistic corrections to the thermal Sunyaev-Zel'dovich effect in the form of the Fokker-Planck expansion.  In their derivation, the scheme to conserve the photon number at every stage of the expansion which has been proposed by Challinor \& Lasenby (1998) played an essential role.  The results of Challinor \& Lasenby (1998) are in agreement with those of Itoh, Kohyama, \& Nozawa (1998).  The latter results include higher-order expansions.  Itoh, Kohyama, \& Nozawa (1998) have also calculated the collision integral of the Boltzmann equation numerically and have compared the results with those obtained by the Fokker-Planck expansion method.  They have confirmed that the Fokker-Planck expansion method gives an excellent result for $k_{B}T_{e} \leq 15$keV, where $T_{e}$ is the electron temperature.  For $k_{B}T_{e} \geq 15$keV, however, the Fokker-Planck expansion results show nonnegligible deviations from the results obtained by the numerical integration of the collision term of the Boltzmann equation.  Here it should be pointed out that the generalized Kompaneets equation is equivalent to a single-scattering approximation.  Thus for high-temperature clusters 
($k_{B}T_{e} \geq $15 keV) the relativistic corrections may underestimate the Sunyaev-Zel'dovich effect at high frequencies.

  Nozawa, Itoh, \& Kohyama (1998b) have extended their method to the case where the galaxy cluster is moving with a peculiar velocity with respect to CMB.  They have thereby obtained the relativistic corrections to the kinematical Sunyaev-Zel'dovich effect.  Challinor \& Lasenby (1999) have confirmed the correctness of the result obtained by Nozawa, Itoh, \& Kohyama (1998b).  Sazonov \& Sunyaev (1998a, b) have calculated the kinematical Sunyaev-Zel'dovich effect by a different method.  Their results are in agreement with those of Nozawa, Itoh, \& Kohyama (1998b).  The latter authors have given the results of the higher-order expansions.

  Itoh, Nozawa, \& Kohyama (2000) have also applied their method to the calculation of the relativistic corrections to the polarization Sunyaev-Zel'dovich effect (Sunyaev \& Zel'dovich 1980b, 1981).  They have thereby confirmed the result of Challinor, Ford, \& Lasenby (1999) which has been obtained with a completely different method.  Recent works on the polarization Sunyaev-Zel'dovich effect include Audit \& Simons (1998), Hansen \& Lilje (1999), and Sazonov \& Sunyaev (1999).

  In the present paper we address ourselves to the numerical calculation of the relativistic corrections to the thermal Sunyaev-Zel'dovich effect.  As stated above, Itoh, Kohyama, \& Nozawa (1998) have carried out the numerical integration of the collision term of the Boltzmann equation.  This method produces the exact results without the power series expansion approximation.  In view of the recent discovery of an extremely high temperature galaxy cluster with $k_{B}T_{e} = 17.4 \pm 2.5$keV (Tucker et al. 1998), it would be extremely useful to present the results of the numerical integration of the collision term of the Boltzmann equation in the form of an accurate analytic fitting formula.

  Sazonov \& Sunyaev (1998a, b) have reported the results of the Monte Carlo calculations on the relativistic corrections to the Sunyaev-Zel'dovich effect.  In Sazonov \& Sunyaev (1998b), a numerical table which summarizes the results of the Monte Carlo calculations has been presented.  This table is of great value when one wishes to calculate the relativistic corrections to the Sunyaev-Zel'dovich effect for galaxy clusters of extremely high temperatures.  Accurate analytic fitting formulae would be still more convenient to use for the observers who wish to analyze the galaxy clusters with extremely high temperatures.  This is the motivation of the present paper.  For the analyses of the galaxy clusters with extremely high temperatures, the results of the calculation of the relativistic thermal bremsstrahlung Gaunt factor (Nozawa, Itoh, \& Kohyama 1998a) and their accurate analytic fitting formulae (Itoh et al. 2000) will be useful.

  The present paper is organized as follows.  In $\S$ 2 we give the method of the calculation.  In $\S$ 3 we give the analytic fitting formula.  Concluding remarks will be given in $\S$ 4.

\section{BOLTZMANN EQUATION}

  We will formulate the kinetic equation for the photon distribution function using a relativistically covariant formalism (Berestetskii, Lifshitz, \& Pitaevskii 1982; Buchler \& Yueh 1976).  As a reference system, we choose the system which is fixed to the center of mass of the cluster of galaxies.  This choice of the reference system affords us to carry out all the calculations in the most straightforward way.  We will use the invariant amplitude for the Compton scattering as given by Berestetskii, Lifshitz, \& Pitaevskii (1982) and by Buchler \& Yueh (1976).

 The time evolution of the photon distribution function $n(\omega)$ is written as 
\begin{eqnarray}
\frac{\partial n(\omega)}{\partial t} & = & -2 \int \frac{d^{3}p}{(2\pi)^{3}} d^{3}p^{\prime} d^{3}k^{\prime} \, W \,
\left\{ n(\omega)[1 + n(\omega^{\prime})] f(E) - n(\omega^{\prime})[1 + n(\omega)] f(E^{\prime}) \right\} \, ,  \\
W & = & \frac{(e^{2}/4\pi)^{2} \, \overline{X} \, \delta^{4}(p+k-p^{\prime}-k^{\prime})}{2 \omega \omega^{\prime} E E^{\prime}} \, ,  \\
\overline{X} & = & - \left( \frac{\kappa}{\kappa^{\prime}} + \frac{\kappa^{\prime}}{\kappa} \right) + 4 m^{4} \left( \frac{1}{\kappa} + \frac{1}{\kappa^{\prime}} \right)^{2} 
 - 4 m^{2} \left( \frac{1}{\kappa} + \frac{1}{\kappa^{\prime}} \right) \, ,  \\
\kappa & = & - 2 (p \cdot k) \, = \, - 2 \omega E \left( 1 - \frac{\mid \vec{p} \mid}{E} {\rm cos} \alpha \right) \, ,  \\
\kappa^{\prime} & = &  2 (p \cdot k^{\prime}) \, = \, 2 \omega^{\prime} E \left( 1 - \frac{\mid \vec{p} \mid}{E} {\rm cos} \alpha^{\prime} \right) \, .
\end{eqnarray}
In the above $W$ is the transition probability corresponding to the Compton scattering.  The four-momenta of the initial electron and photon are $p = (E, \vec{p})$ and $k = (\omega, \vec{k})$, respectively.  The four-momenta of the final electron and photon are $p^{\prime} = (E^{\prime}, \vec{p}^{\prime})$ and $k^{\prime} = (\omega^{\prime}, \vec{k}^{\prime})$, respectively.  The angles $\alpha$ and $\alpha^{\prime}$ are the angles between $\vec{p}$ and $\vec{k}$, and between $\vec{p}$ and $\vec{k}^{\prime}$, respectively.  Throughout this paper, we use the natural unit $\hbar = c = 1$ unit, unless otherwise stated explicitly.

  By ignoring the degeneracy effects, we have the relativistic Maxwellian distribution for electrons with temperature $T_{e}$ as follows
\begin{eqnarray}
f(E) & = & \left[ e^{\left\{(E - m)-(\mu - m) \right\}/k_{B}T_{e}} \, + \, 1 \right]^{-1}  \nonumber \\
& \approx & e^{-\left\{K-(\mu - m)\right\}/k_{B}T_{e}} \, ,
\end{eqnarray}
where $K \equiv (E - m)$ is the kinetic energy of the initial electron, and $(\mu - m)$ is the non-relativistic chemical potential of the electron. 
We now introduce the quantities
\begin{eqnarray}
x &  \equiv &  \frac{\omega}{k_{B}T_{e}}  \, ,  \\
\Delta x &  \equiv &  \frac{\omega^{\prime} - \omega}{k_{B}T_{e}}  \, .
\end{eqnarray}
Substituting equations (2.6) -- (2.8) into equation (2.1), we obtain
\begin{equation}
\frac{\partial n(\omega)}{\partial t} =  -2 \int \frac{d^{3}p}{(2\pi)^{3}} d^{3}p^{\prime} d^{3}k^{\prime} \, W \, f(E) \,
\left\{ [1 + n(\omega^{\prime})] n(\omega) -  [1 + n(\omega)] n(\omega^{\prime}) e^{ \Delta x }  \right\} \, .
\end{equation}
Equation (2.9) is our basic equation.  We will denote the Thomson scattering cross section by $\sigma_{T}$, and the electron number density by $N_{e}$.  We will define
\begin{eqnarray}
\theta_{e} & \equiv & \frac{k_{B}T_{e}}{m_{e}c^{2}}  \, ,  \\
y & \equiv & \sigma_{T} \int d \ell N_{e} \, ,
\end{eqnarray}
where $T_{e}$ is the electron temperature, and the integral in equation (2.11) is over the path length of the galaxy cluster.  By introducing the initial photon distribution of the CMB radiation which is assumed to be Planckian with temperature $T_{0}$
\begin{eqnarray}
n_{0} (X) & = & \frac{1}{e^{X} - 1} \, ,   \\
X & \equiv & \frac{\omega}{k_{B} T_{0}}  \, ,
\end{eqnarray}
we rewrite equation (2.9) as
\begin{equation}
\frac{\Delta n(X)}{n_{0}(X)} \, = \, y \, F(\theta_{e}, X)  \, .
\end{equation}
We obtain the function $F(\theta_{e}, X)$ by numerical integration of the collision term of the Boltzmann equation (2.9).  The accuracy of the numerical integration is about $10^{-5}$.  We confirm that the condition of the photon number conservation
\begin{equation}
\int d X \, X^{2} \, \Delta n(X) \, = \, 0  \,
\end{equation}
is satisfied with the accuracy better than $10^{-9}$.

  We define the distortion of the spectral intensity as
\begin{eqnarray}
\Delta I & \equiv & \frac{X^{3}}{e^{X} - 1} \, \frac{\Delta n(X)}{n_{0}(X)}  \,  \\
& = &  y \, \frac{X^{3}}{e^{X} - 1} \, F(\theta_{e}, X)  \, .
\end{eqnarray}
The graph of $F(\theta_{e},X)$ is shown in Figure 1.  The graph of $\Delta I/y$ is shown in Figure 2.

\section{ANALYTIC FITTING FORMULA}

  We give an accurate analytic fitting formula for the function $F(\theta_{e}, X)$ in equation (2.14) which has been obtained by numerical integration of the collision term of the Boltzmann equation.  We will give an analytic fitting formula for the ranges $0.02 \leq \theta_{e} \leq 0.05$, $0 \leq X \leq 20$, which will be sufficient for the analyses of the galaxy clusters.  For $\theta_{e} < 0.02$, the results of Itoh, Kohyama, and Nozawa (1998) give sufficiently accurate results (the accuracy is generally better than 1\%).

  We express the fitting formula for $0.02 \leq \theta_{e} \leq 0.05$ as follows:
\begin{eqnarray}
\frac{\Delta n(X)}{n_{0}(X)} & = & y \, F(\theta_{e}, X)   \nonumber  \\
& = &  y \, \left\{ \frac{\theta_{e} X e^{X}}{e^{X}-1} \, \left(
Y_{0} \, + \, \theta_{e} Y_{1} \, + \, \theta_{e}^{2} Y_{2} \, + \,  \theta_{e}^{3} Y_{3} \, + \,  \theta_{e}^{4} Y_{4} \, \right) \, + R  \right\}  \, .
\end{eqnarray}
The functions $Y_{0}$, $Y_{1}$, $Y_{2}$, $Y_{3}$, and $Y_{4}$ have been obtained by Itoh, Kohyama, and Nozawa (1998) with the Fokker-Planck expansion method, and their explicit expressions have been given.  

  We define the residual function $R$ in equation (3.1) as follows:
\begin{eqnarray}
R & = & \left\{ \begin{array}{ll}  0 \, ,  &  \mbox{for $0 \leq X < 2.5$}  \\
        \displaystyle{ \sum_{i,j=0}^{10} \, a_{i \, j} \, \Theta_{e}^{i} \, Z^{j} } \, ,  & 
      \mbox{for $2.5 \leq X \leq 20.0$  \, , }
         \end{array}    \right.
\end{eqnarray}
where
\begin{eqnarray}
\Theta_{e} & \equiv & 25 \left( \theta_{e} \, - \, 0.01 \right)  \, \, , \, \, \, \, \, \, 0.02 \leq \theta_{e} \leq 0.05  \, ,  \\
Z  & \equiv &  \frac{1}{17.6} \left( X \, - \, 2.4 \right)  \, \, , \, \, \, \, \, \, 2.5 \leq  X  \leq  20.0 \, .
\end{eqnarray}
The coefficients $a_{i \, j}$ are presented in TABLE 1.  The accuracy of the fitting formula for equation (3.1) is generally better than 0.1\% except for a region of
$\theta_{e}=0.05$, $X>17$, where the error exceeds 1\%.

\section{CONCLUDING REMARKS}

We have calculated the relativistic corrections to the thermal Sunyaev-Zel'dovich effect for clusters of galaxies by numerical integration of the collision term of the Boltzmann equation.  We have presented an accurate analytic fitting formula for the thermal Sunyaev-Zel'dovich effect.  The fitting formula covers all the ranges of the observation of galaxy clusters in the foreseeable future.  The accuracy of the fitting is generally better than 0.1\%.  The present results will be useful for the analyses of the galaxy clusters with extremely high temperatures.  For galaxy clusters with relatively low temperatures $\theta_{e} < 0.02$, the Fokker-Planck expansion results of Itoh, Kohyama, \& Nozawa (1998) will be sufficiently accurate (the accuracy is generally better than 1\%).

\acknowledgements

  We thank Professor Y. Oyanagi for allowing us to use the least square fitting program SALS.  We also thank our anonymous referee for many valuable comments which helped us tremendously in revising the manuscript.  This work is financially supported in part by the Grant-in-Aid of Japanese Ministry of Education, Science, Sports, and Culture under the contract \#10640289.

\newpage


\references{} 
\reference{} Arnaud, K. A., Mushotzky, R. F., Ezawa, H., Fukazawa, Y., Ohashi, T., Bautz, M. W., Crewe, G. B., Gendreau, K. C., Yamashita, K., Kamata, Y., \& Akimoto, F. 1994, ApJ, 436, L67
\reference{} Audit, E., \& Simmons, J. F. L. 1999, MNRAS, 305, L27
\reference{} Berestetskii, V. B., Lifshitz, E. M., \& Pitaevskii, L. P. 1982, $Quantum$ $Electrodynamics$ (Oxford: Pergamon)
\reference{} Birkinshaw, M. 1979, MNRAS, 187, 847
\reference{} Birkinshaw, M., \& Hughes, J., P. 1994, ApJ, 420, 33
\reference{} Birkinshaw, M., Hughes, J. P., \& Arnaud, K. A. 1991, ApJ, 379, 466
\reference{} Buchler, J. R., \& Yueh, W. R. 1976, ApJ, 210, 440
\reference{} Cavaliere, A., Danese, L., \& De Zotti, G. 1979, A\&A, 75, 322
\reference{} Challinor, A., Ford, M., \& Lasenby, A., 1999, MNRAS in press
\reference{} Challinor, A., \& Lasenby, A., 1998, ApJ, 499, 1
\reference{} Challinor, A., \& Lasenby, A., 1999, ApJ, 510, 930
\reference{} David, L. P., Slyz, A., Jones, C., Forman, W., \& Vrtilek, S. D. 1993, ApJ, 412, 479
\reference{} Furuzawa, A., Tawara, Y., Kunieda, H., Yamashita, K., Sonobe, T., Tanaka, Y., \& Mushotzky, R. 1998, ApJ, 504, 35
\reference{} Gunn, J. E. 1978, in Observational Cosmology, 1, ed. A. Maeder, L. Martinet \& G. Tammann (Sauverny: Geneva Obs.)
\reference{} Hansen, E., \& Lilje, P. B. 1999, MNRAS, 306, 153
\reference{} Herbig, T., Lawrence, C. R., Readhead, A. C. S., \& Gulkus, S. 1995, ApJ, 449, L5
\reference{} Holzapfel, W. L. et al. 1997, ApJ, 480, 449
\reference{} Itoh, N., Kohyama, Y., \& Nozawa, S. 1998, ApJ, 502, 7
\reference{} Itoh, N., Nozawa, S., \& Kohyama, Y. 2000, ApJ in press
\reference{} Itoh, N., Sakamoto, T., Kusano, S., Nozawa, S., \& Kohyama, Y. 2000, ApJS in press
\reference{} Jones, M. 1995, Astrophys. Lett. Commun., 6, 347
\reference{} Kompaneets, A. S. 1957, Soviet Physics JETP, 4, 730
\reference{} Markevitch, M. 1998, ApJ, 504, 27
\reference{} Markevitch, M., Mushotzky, R., Inoue, H., Yamashita, K., Furuzawa, A., \& Tawara, Y. 1996, ApJ, 456, 437
\reference{} Markevitch, M., Yamashita, K., Furuzawa, A., \& Tawara, Y. 1994, ApJ, 436, L71
\reference{} Myers, S. T., Baker, J. E., Readhead, A. C. S., \& Herbig, T. 1995, preprint
\reference{} Mushotzky, R. F., \& Scharf, C. A. 1997, ApJ, 482, L13
\reference{} Nozawa, S., Itoh, N., \& Kohyama, Y. 1998a, ApJ, 507, 530
\reference{} Nozawa, S., Itoh, N., \& Kohyama, Y. 1998b, ApJ, 508, 17
\reference{} Rephaeli, Y. 1995, ApJ, 445, 33
\reference{} Rephaeli. Y., \& Yankovitch, D. 1997, ApJ, 481, L55
\reference{} Sazonov, S. Y., \& Sunyaev, R. A. 1998a, ApJ, 508, 1
\reference{} Sazonov, S. Y., \& Sunyaev, R. A. 1998b, Astronomy Letters 24, 553
\reference{} Sazonov, S. Y., \& Sunyaev, R. A. 1999, MNRAS in press
\reference{} Silk, J. I., \& White, S. D. M. 1978, ApJ, 226, L103
\reference{} Stebbins, A., 1997, preprint astro-ph/9705178
\reference{} Sunyaev, R. A., \& Zel'dovich, Ya. B. 1972, Comments Astrophys. Space Sci., 4, 173
\reference{} Sunyaev, R. A., \& Zel'dovich, Ya. B. 1980a, ARA\&A, 18, 537
\reference{} Sunyaev, R. A., \& Zel'dovich, Ya. B. 1980b, MNRAS, 190, 413
\reference{} Sunyaev, R. A., \& Zel'dovich, Ya. B. 1981, Astrophysics and Space Physics Reviews, 1, 1
\reference{} Tucker, W., Blanco, P., Rappoport, S., David, L., Fabricant, D., Falco, E. E., Forman, W., Dressler, A., \& Ramella, M. 1998, ApJ, 496, L5
\reference{} Weymann, R. 1965, Phys. Fluid, 8, 2112
\reference{} Zel'dovich, Ya. B., \& Sunyaev, R. A. 1969, Astrophys. Space Sci., 4, 301


\newpage
\centerline{\bf \large Figure Captions}

\begin{itemize}

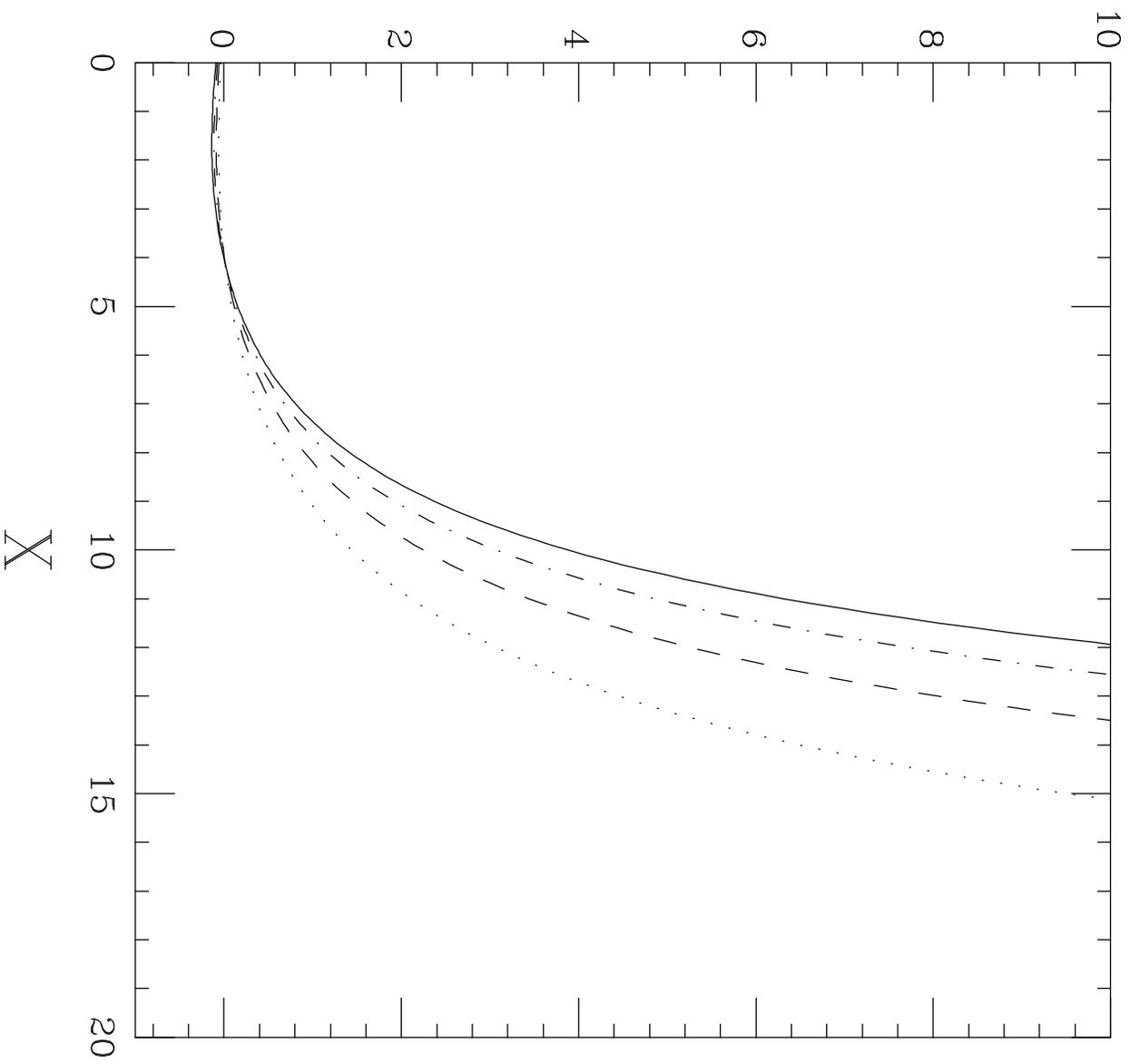
\item Fig.1. The function $F(\theta_{e}, X)$.  The dotted, dashed, dash-dotted, and solid curves correspond to the cases for $\theta_{e}=0.02, 0.03, 0.04, 0.05$, respectively.

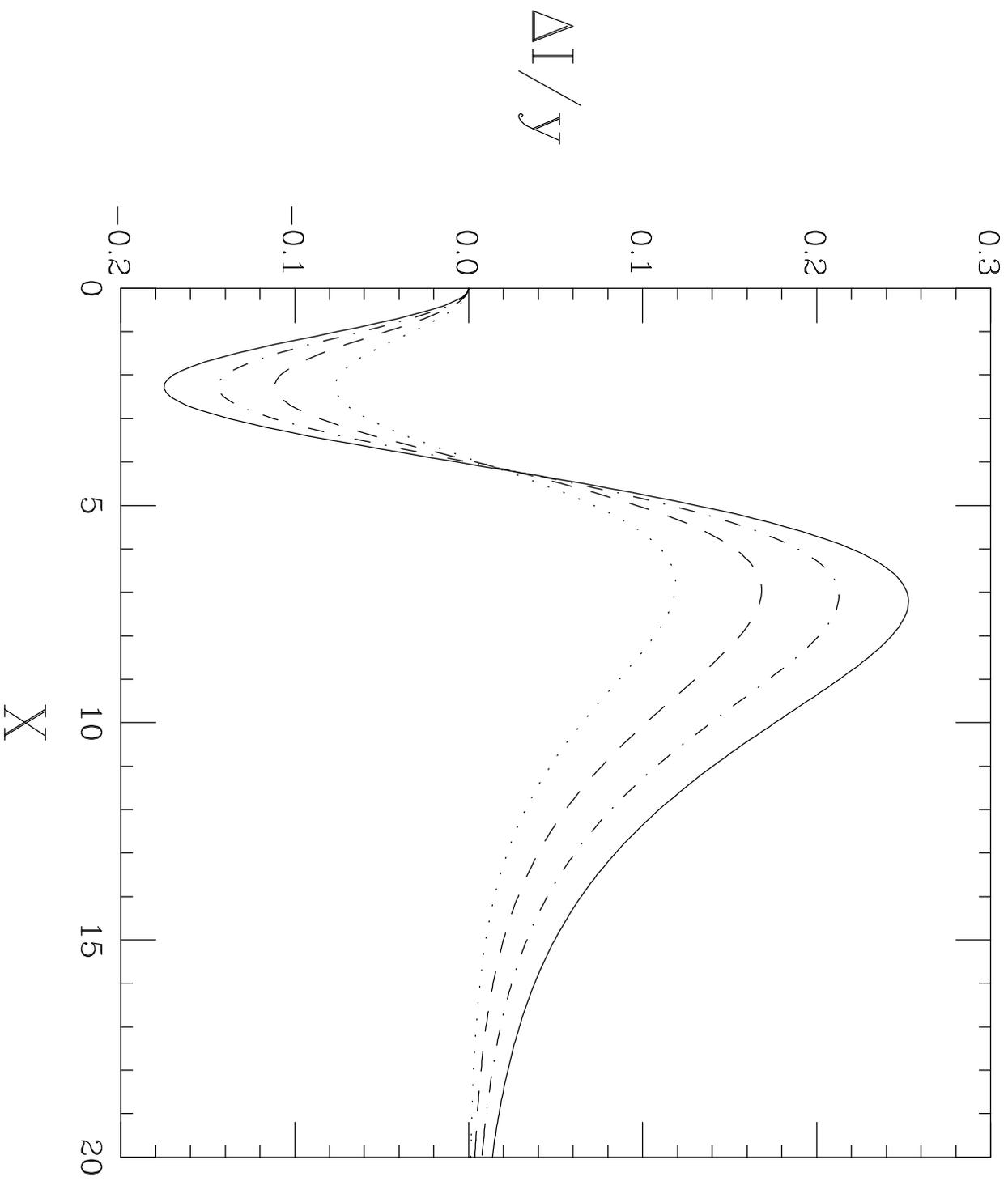
\item Fig.2. The spectral intensity distortion $\Delta I/y$ as a function of $X$.  The dotted, dashed, dash-dotted, and solid curves correspond to the cases for $\theta_{e}=0.02, 0.03, 0.04, 0.05$, respectively.

\end{itemize}

\end{document}